\documentclass[12pt]{article}

\usepackage[english]{babel}
\usepackage{natbib}
\usepackage{booktabs} 
\usepackage[letterpaper,top=2cm,bottom=2cm,left=3cm,right=3cm,marginparwidth=1.75cm]{geometry}

\usepackage{amsmath}
\usepackage{graphicx}
\usepackage{setspace}
\usepackage[colorlinks=true, allcolors=blue]{hyperref}
\usepackage{placeins} 
\usepackage{subcaption} 

\title{Integrating Large Language Models and Reinforcement Learning for Sentiment-Driven Quantitative Trading}
\author{Wo Long, Wenxin Zeng, Xiaoyu Zhang, Ziyao Zhou}
\begin{document}
\maketitle

\begin{abstract}
This research develops a sentiment-driven quantitative trading system that leverages a large language model —FinGPT—for sentiment analysis, and explores a novel method for signal integration using a reinforcement learning algorithm, Twin Delayed Deep Deterministic Policy Gradient (TD3). We compare the performance of strategies that integrate sentiment and technical signals using both a conventional rule-based approach and a reinforcement learning framework. The results suggest that sentiment signals generated by FinGPT offer value when combined with traditional technical indicators, and that reinforcement learning algorithm presents a promising approach for effectively integrating heterogeneous signals in dynamic trading environments.
\end{abstract}

\section{Introduction}
The increasing availability of unstructured data has opened new frontiers in quantitative finance. In particular, the integration of sentiment analysis into trading strategies has gained great interest. In contrast to traditional technical indicators, which capture patterns in historical price and volume data, sentiment signals extracted from news articles and other media offer a complementary, forward-looking perspective rooted in investor expectations and market narratives. However, effectively combining these two distinct sources of information, one backward-looking and one anticipatory, remains a significant challenge in systematic investing.
\newline

This paper explores an innovative approach to integrating sentiment information with traditional technical indicators in equity market trading. We propose a framework driven by reinforcement learning to dynamically combine these two categories of signals and evaluate the performance of portfolios constructed based on this method. To provide a benchmark, we also compare the RL-based integration with a conventional rule-based strategy. The rule-based method constructs a composite trading signal by linearly weighting standardized sentiment and technical scores, offering an intuitive and easily adjustable framework. In contrast, the RL agent learns to synthesize and act on these signals through continuous interaction with the market environment, optimizing for long-term and risk-adjusted returns. This comparison enables us to assess the added value of reinforcement learning in the blending of heterogeneous information for portfolio decision making.
\newline

The paper is guided by three main objectives. First, his research aims to examine whether sentiment signals extracted from financial news and generated by a large language model have significant predictive power over stock returns. Second, if the sentiment signals generated can enhance the traditional trading strategies based on technical indicators. The third and most innovative contribution of this paper is to evaluate whether reinforcement learning provides a valid and effective approach for integrating sentiment and technical signals into a better-performing trading strategy.
\newline

The paper is organized as follows. Section 2 reviews the existing literature on language models, reinforcement learning, and related areas, and discusses the potential contributions of this study. Section 3 describes the data used in this study along with the pre-processing pipeline. Section 4 presents the technical framework of the trading system, detailing the specific language models and reinforcement learning algorithms used, and explaining how they interact dynamically to generate trading decisions. Section 5 outlines the detailed implementation of the trading strategies, including the trading rules and execution assumptions for both the traditional rule-based approach and the reinforcement learning–driven strategy. Section 6 and 7 discusses the results and further implication.

\section{Literature Review and Contribution}

The integration of Large Language Models into financial decision making has seen significant advancement over the past years, including sectors such as quantitative trading, credit scoring, and even compliance and regulation. In quantitative trading, the application of LLMs can be roughly broken down into categories based on two types of data: text-based application and figures-based application. Text data are often leveraged for sentiment analysis, which has now become more and more crucial in stock picking, timing, and prediction. \cite{bernard2023modular} use a GPT large language model fine-tuned on narrative disclosures and inline XBRL tags to predict what numbers represent based on surrounding text. \cite{LopezLira2023} document the capability of LLMs to predict stock price movements using news headlines without direct financial training of the models. \cite{glasserman2023new} quantify news novelty, which refers to the changes in the distribution of news text, through an entropy measure. The entropy-based trading strategy generated statistically significant abnormal returns and alpha. \cite{zhou2025end} introduce an end-to-end trading system that leverages LLMs for real-time market sentiment analysis by synthesizing data from financial news and social media.
\newline

Numerical data has also gained increasing attention in quantitative investing areas. Recent studies have also brought attention to LLMs’ potential application in analyzing figures in financial statements to calculate a company’s future earnings direction. \cite{kim2024financial} tested the predictive power of LLMs on earning's direction using a narrow information set that includes numerical information reported on two financial statements, \textit{i.e.}, balance sheet and income statement. The study also stated that their results were to be the “lower bound” of LLMs’ predictive power since only numerical information was incorporated, while textual analysis was the main strength of large language models and most aligned with LLMs’ initial purpose. This marks a milestone in LLMs’ application in finance and quantitative investment because it has discovered LLMs’ potential to outperform fundamental analysts, who make predictions based on a comprehensive understanding of accounting statements.
\newline

While these applications highlight the growing role of LLMs in finance, \cite{sarkar2024lookahead} raise concerns about look-ahead bias, where pretraining data may unintentionally includes future information, potentially contaminating predictive analyses. Through empirical tests, they demonstrate that LLMs can generate future-dependent insights, such as predicting post-2019 risks (\textit{e.g.}, the COVID-19 pandemic) from earnings calls prior to 2020. They also show that LLMs achieve a 70–80\% accuracy in predicting close election outcomes, despite such events being considered unpredictable ex-ante. Their findings emphasize the importance of carefully managing pretraining data cutoffs to ensure robust and reliable financial forecasting. A common approach in prior literature to mitigate look-ahead bias involves using a consistently anonymized format for financial texts, in which specific entities are masked—making it virtually impossible for the model to infer a firm’s identity based on the text structure (\cite{kim2024financial}, \cite{glasserman2023assessing}). This method has been empirically shown by \cite{kim2024financial} to be effective in structured texts such as financial statements, where the content follows a uniform and predictable format. \cite{kim2024financial} also implemented a robustness check to assess potential information leakage in LLMs by evaluating the model's ability to guess firm-specific entities from masked texts.
\newline

Reinforcement learning has emerged as a powerful approach for data-driven decision-making in trading and portfolio allocation, allowing models to dynamically adjust strategies based on evolving market conditions. FinRL \cite{liu2020finrl} introduces a deep reinforcement learning framework that automates stock trading by training agents using real market environments while incorporating trading constraints such as transaction costs and risk-aversion levels. The FinRL library supports various reinforcement learning algorithms and provides standardized training, validation, and back-testing pipelines. Building on this, FinRL-Meta \cite{liu2022finrl} improves market simulation quality by introducing hundreds of gym-style market environments with dynamic data updates, addressing common challenges such as low signal-to-noise ratio, survivorship bias, and model overfitting. These frameworks demonstrate that RL-based trading agents can outperform traditional strategies by optimizing risk-adjusted returns and improving execution efficiency. However, while reinforcement learning has shown promise, the challenge remains in ensuring robust generalization to unseen market conditions and mitigating overfitting to historical data.
\newline

Despite the expanding body of research on large language models in finance, limited attention has been given to integrating sentiment information with traditional technical signals. This paper contributes to the literature in several ways. First, it introduces sentiment signals generated by a state-of-the-art language model—FinGPT. The long-short strategy based on these sentiment signals exhibits returns that are not explained by conventional factors such as market, size, and value. Second, this paper evaluates whether reinforcement learning algorithms offer a robust framework for translating LLM-generated sentiment signals—when combined with technical indicators—into actionable trades and portfolio allocations. Specifically, it extends the work of \cite{zhou2025end} by refining LLM-based signals through an RL-driven decision-making pipeline, and compares the results against traditional buy-and-hold benchmarks. Third, this paper addresses and mitigates the look-ahead bias that often arises in LLM-related research. By incorporating stricter data pre-processing techniques and performing robustness checks, similar to those used in \cite{kim2024financial}, the study enhances the reliability of the resulting trading strategies. Finally, this paper compares model-free reinforcement learning strategies with conventional rule-based trading approaches, evaluating their effectiveness in dynamic and volatile market environments.

\section{Data}

\subsection{Data Description}
The primary data used in this research include news articles data and stock price and volume data. We use a universe of 44 stocks selected in S\&P 500 from 2018 to 2025, setting aside data from 2024 to 2025 for back-testing and evaluating the out-of-sample performance of the proposed strategies. The news data are sourced from Thomson Reuters, covering the period from 2018 to 2025, and include the 44 S\&P 500 stocks selected based on the most active level of news coverage. The 44 companies in the selected universe exhibit relatively consistent news coverage throughout the 2018 to 2025 period. This universe of stocks is intentionally chosen as the objective of this research is to assess the predictive power and incremental value of sentiment information on trading performance, making it logical to concentrate on companies with consistently rich new flow. Each news observation is associated with a specific company identifier and a trading day timestamp. Multiple news articles often available for a single firm on a given trading day. We also dropped news articles released after 4 p.m. on each trading day to ensure the viability of sentiment signals for same-day trading decisions.
\newline

The historical price and volume data for the selected universe are daily-level data sourced from CRSP and Bloomberg, spanning also from 2018 to 2025. The technical data include primarily open and close prices, trading volume (measured as number of shares traded per day) obtained from CRSP, and VWAP sourced from Bloomberg. 

\subsection{Data Processing}

The processing of news articles is a critical component of the research. Financial news tends to be lengthy and often contains redundant or irrelevant information, which might jeopardize the accuracy of sentiment classification and substantially increase the computational costs. To enable more efficient analysis, we implement a summarization step. Specifically, we employ LLaMA 3.1 8B to condense raw news articles into concise, company-level daily summaries. As mentioned earlier, multiple news articles are often available for a single firm on a given trading day. After the summarization step, all related articles for a company are condensed into a single summary per trading day. Note that the summarization on trading day $t$ only involves news released before 4 p.m. on the same day. The step leave us with 44 (number of stocks) $\times$ 1848 (trading days from 2018 to January 2025) = 81,312 observations.
\newline

To improve scalability and computational efficiency, we adopted vLLM \cite{kwon2023efficient}, a framework optimized specifically for efficient inference with large language models. By enabling batch processing of queries, vLLM considerably boosts the speed and throughput of our summarization tasks, especially when handling extensive financial news datasets. We integrated vLLM with our LLaMA model to achieve a balance of efficiency and high-quality summaries. During inference, texts are broken into manageable segments to effectively handle large volumes of data while preserving the context necessary for accurate summarization. Sampling methods are thoughtfully adjusted to ensure summaries remain consistent, precise, and focused, resulting in concise and meaningful insights.
\newline

Price and volume data are adjusted for dividends, stock splits, and other corporate actions to ensure consistency across time. Price, volume and news data are synchronized at the daily level to generate coherent and meaningful signals for each trading day.

\section{Technical Framework}

\subsection{Overview}
The trading system illustrated in Figure 1 leverages cutting-edge large language models and reinforcement learning algorithms to generate trading decisions. This architecture integrates real-time sentiment analysis with historical technical indicators, forming a comprehensive strategy that incorporates both sentiment information and price-volume dynamics. 
\newline

\begin{figure}[h]
\centering
\includegraphics[width=0.9\linewidth]{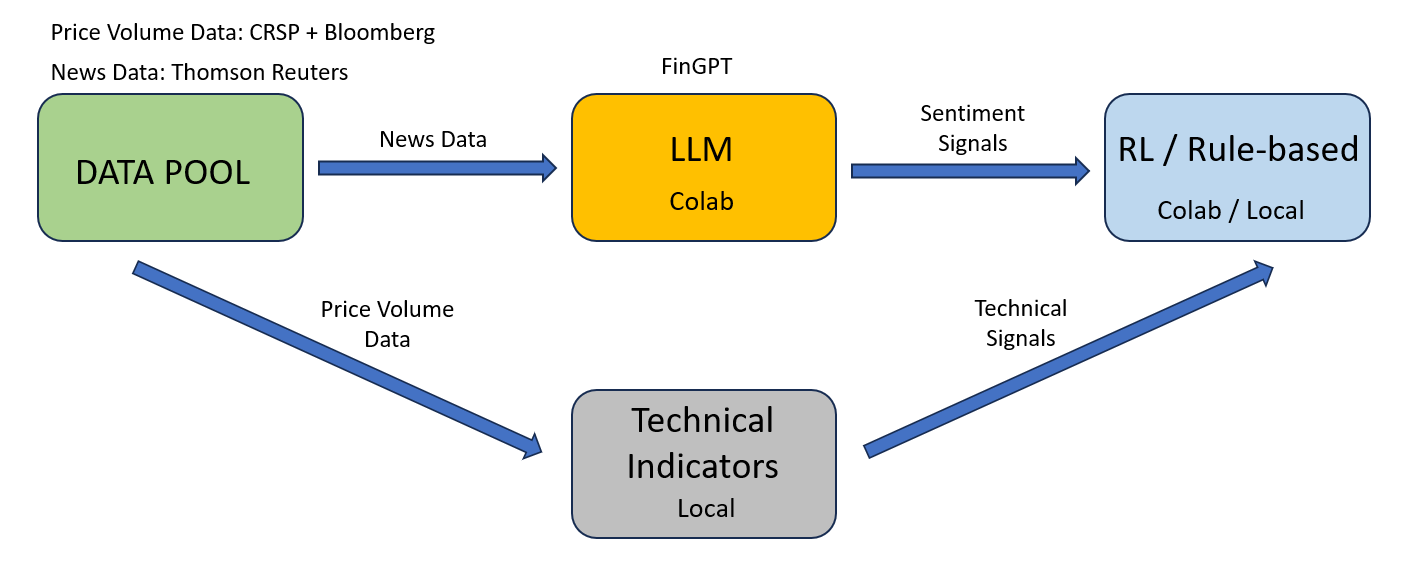}
\caption{\label{fig:illustation} Trading System Illustration}
\end{figure}

First, news data are processed using a large language model (\textit{i.e.}, FinGPT), which is trained on the Google Colab platform to generate sentiment signals. Simultaneously, a range of technical indicators, including RSI (Relative Strength Index), VWAP (Volume-Weighted Average Price), MACD (Moving Average Convergence Divergence), \textit{etc.}, is calculated based on historical price and volume data. These sentiment and technical signals are then used as inputs for both reinforcement learning–based and conventional rule-based strategies to generate buy and sell decisions.

\subsection{Large Language Model: FinGPT}

Although large language models have demonstrated strong capabilities in understanding and evaluating sentiment in textual data, they often underperform in financial applications—particularly in volatile, short-term markets—due to the substantial differences between general-purpose language and domain-specific financial text. \cite{wu2023bloomberggptlargelanguagemodel} introduced BloombergGPT, a powerful financial large language model trained on a vast range of financial data, which excels in tasks such as named entity recognition and sentiment analysis (\cite{zhou2025end}). However, the model is closed-source—implying high costs for implementation—and lacks empirical validation in trading applications. In light of these challenges, \cite{liu2023fingptdemocratizinginternetscaledata} introduced FinGPT, an open-source, domain-specific large language model pretrained on financial texts, which also offers practitioners simple yet effective strategies for fine-tuning on domain-specific financial data. \cite{zhou2025end} tested the FinGPT model in live trading environments and demonstrated its strong performance in accurately capturing nuanced market language and translating it into effective trading decisions.
\newline

This paper uses a version of the FinGPT model that was pre-trained on data available up to November 2023. The company-level daily news summaries are processed by FinGPT for sentiment classification. Each summary piece is categorized into one of three sentiment classes: [Positive, Neutral, Negative], along with a confidence score (logit). The confidence score is a key input to both rule-based and RL-driven trading strategies. Sentiment signal $sentiment_{i,t}$ for company $i$ on trading day $t$ is generated using the summarized news of company $i$ on trading day $t$. 

\subsection{Technical Indicators}
Technical signals are computed using price and volume data. These include:

\begin{itemize}
    \item Relative Strength Index (RSI): Measures the speed and change of price movements, indicating overbought or oversold conditions.
    \item Volume Weighted Average Price (VWAP): Reflects the average price weighted by volume, useful for intra-day trading analysis and price benchmarking.
    \item Moving Average Convergence Divergence (MACD): Highlights changes in the strength, direction, and momentum of a trend, assisting in identifying potential buy or sell signals.
    \item Garman-Klass volatility: Estimator for the true volatility of a financial asset over a given period. It uses high, low, open, and close prices to provide a more efficient estimate than the traditional close-to-close volatility.
\end{itemize}

All features were z-score normalized and used to construct a technical alpha signal. Technical signals $technical_{i,t}$ for company $i$ on trading day $t$ is generated using the price and volume data of company $i$ from trading day $t-1$.

\subsection{Reinforcement Learning Algorithm}
\label{subsec:rl_algo}

We adopt \textbf{ twin delayed deep deterministic policy gradient} (TD3), a model-free, off-policy actor–critic algorithm designed for continuous action control. TD3 extends the deep deterministic policy gradient (DDPG) to reduce the two principal failure modes of deterministic RL (\textit{i.e.}, over-estimation bias and high target variance) through three modifications:

\begin{enumerate}
  \item \textbf{Twin critics.} Two independent Q-networks $Q_{\theta_1},Q_{\theta_2}$ are trained and the minimum of their target estimates is used when boot-strapping:
        \[
          y_t = r_t + \gamma \,
          \min_{j=1,2} Q_{\theta'_j}\!\bigl(s_{t+1},\,\pi_{\phi'}(s_{t+1})+\epsilon\bigr).
        \]
        This simple change halves the over-estimation in practice.
        
  \item \textbf{Target-policy smoothing.} Small clipped Gaussian noise $\epsilon\sim\mathcal N(0,\sigma^2)$ is added to the next-state action so that the critic learns a value integrated over a region of the action space, yielding smoother gradients and improving robustness of noisy market rewards.
  \item \textbf{Delayed policy updates.} The actor $\pi_\phi$ is updated less frequently, that is, every $d=2$ critic steps, giving the critics time to converge and producing more stable policy learning.
\end{enumerate}

The TD3 algorithm is suitable for portfolio allocation primarily for three reasons. First, it supports continuous action spaces, which handles naturally the structures of assets weights, whether constrained to a simplex (\textit{i.e.}, long-only portfolio) or hyper-cube (\textit{i.e.}, long-short strategies with leverage constraints). Second, TD3 is well-equipped to deal with the inherent noise in financial data. Financial rewards often exhibit a low signal-to-noise ratio. TD3's conservative value estimation helps reduce bias caused by this characteristic. Third, the algorithm is sample-efficient, as it employs off-policy learning with a replay buffer, which allows the model to repeatedly learn from historical market transitions. This feature is particularly critical when training dataset is limited, with only around 1,800 trading days available.

\paragraph{Network architecture \& hyper-parameters.}
Both actor and critic networks use two fully-connected layers of 256 units with ReLU activations.  
Key training settings:

\begin{center}
\begin{tabular}{@{}lcc@{}}
\toprule
Parameter & Value & Notes \\
\midrule
Learning rate & $10^{-4}$ & Adam optimiser                                 \\
Discount factor $\gamma$ & $0.99$ & Daily compounding                        \\
Replay buffer size & $25\,(T_{\text{train}}-1)$ & $\approx2.0\times10^{5}$  \\
Batch size & $T_{\text{train}}-1$ & One episode-length mini-batch           \\
Policy noise $\sigma$ & $0.2$ & Clipped to \(\pm0.5\)                      \\
Delay $d$ & $2$ & Actor update every 2 critic steps                       \\
Total updates & $5.1\times10^{7}$ & 512 epochs on the 2018–2024 sample      \\
\bottomrule
\end{tabular}
\end{center}

\subsection{Mitigating Look-ahead Bias}
As previously mentioned, the FinGPT model used in this study is pre-trained on data up to November 2023. Therefore, the sentiment scores generated for the out-of-sample testing period from 2024 to 2025 are not subject to look-ahead bias. To further mitigate potential look-ahead bias during sentiment analysis, we follow the methodology in \cite{kim2024financial}, in which all company names, product names, and dates in news articles published before 2024 are masked prior to being processed by FinGPT. 

\section{Methodology and Implementation}

\subsection{Rule-based Trading Strategy}

Standardized technical indicators are aggregated into a \textbf{composite technical alpha}. The final signal used for portfolio construction is a weighted linear combination of the technical and sentiment signals. The technical signal and sentiment signal are constructed as follows:
\begin{equation}
    Technical \ Signal_{i,t} = \dfrac{Volume_{i,t-1}}{MA20_{i,t-1}} + MACD_{i,t-1}
\end{equation}
\begin{equation}
    Sentiment \ Signal_{i,t} = Confidence_{i,t} \cdot I_{\{ 1: Positive; \ -1: Negative;\ 0: Neutral\}}
\end{equation}
Both signals are z-score normalized after construction and then integrated using the following formula:
\begin{equation}
Combined \ Signal_{i,t}= w_t \cdot Technical \ Signal_{i,t} + (1 - w_t) \cdot Sentiment \ Signal_{i,t}
\end{equation}

The default weight is set to \( w_t = 0.5 \), which assigns equal importance to both the technical and sentiment components. This parameter can be adjusted according to market conditions or strategic preferences. The resulting combined score serves as the basis for the construction of the rule-based portfolio, where stocks are ranked daily and sorted into quintiles based on their scores. We use a universe of 44 stocks selected in S\&P 500 from 2018 to 2025, setting aside data from 2024 to 2025 for back-testing and evaluating the out-of-sample performance of the proposed strategies.
\vspace{1em}

\textbf{Execution Assumption:}
\begin{itemize}
    \item Orders are executed at the same day’s closing price (\textit{i.e.}, trades filled at the closing auction (\(\approx\) 4 p.m. ET). )
    \item Long-Short Strategy with initial investment of 0.
    \item We assume zero transaction costs for the rule-based strategy by default. As a robustness check, we also evaluate the strategy’s performance under a transaction cost of 5 basis points per trade.
\end{itemize}

\textbf{Caveats:}
\begin{itemize}
    \item Trades are executed on trading day $t$ using the close price $close_{i,t}$.
\end{itemize}

\textbf{Trading Rules:}
The $Combined\ Signal_{i,t}$ serves as the ranking basis for long-short portfolio construction. Stocks are ranked by the signal each day and assigned to 5 quintiles. The following portfolio longs the top 20\% stocks and shorts the bottom 20\% stocks.

\subsection{RL-Driven Trading Strategy}
\label{subsec:rl_trading}

The feature vector fed to the agent takes into account various factors, including historical returns, price-volume characteristics, volatility, historical assets weights, and most importantly, the sentiment signals generated by FinGPT in previous step. Specifically, observation $s_t$ for trading day $t$ concatenates seven asset-specific characteristics and the previous day portfolio weights:

\begin{itemize}
  \item \textbf{Lagged return} $r_{t-1}$: close-to-close log-return of the previous trading day.
  \item \textbf{Momentum / overbought‐oversold}:
        \begin{itemize}
          \item 14-day relative strength index (RSI\textsubscript{14}).
          \item MACD signal line value (12-26-9 convention).
        \end{itemize}
  \item \textbf{Price-volume microstructure}:
        \begin{itemize}
          \item VWAP gap: $\displaystyle\frac{\text{VWAP}_t}{\text{Close}_t}-1$.
          \item Volume pressure: $\displaystyle \frac{\text{Volume}_t}{\overline{\text{Volume}}_{20}}$.
        \end{itemize}
  \item \textbf{Volatility}:
        \begin{itemize}
          \item Realized volatility ratio: $\displaystyle \frac{\text{RV}_t}{\overline{\text{RV}}_{20}}$,
                where RV is a 5-day close-to-close standard deviation.
          \item Garman–Klass volatility.
        \end{itemize}
  \item \textbf{Sentiment:} $Sentiment\ Signal_{i,t}$ from FinGPT. The signal is processed similarly as in rule-base strategy (\textit{i.e.}, scaled to \([-1,1]\) and multiplied by the confidence scores).
  \item \textbf{Previous portfolio weights} $\mathbf w_{t-1}^{\!\top}$ are also incorporated into the RL framework, allowing the actor to account for transaction costs and portfolio turnover in its decision-making process.
\end{itemize}

\paragraph{Environment design:}
We extend the \texttt{gymnasium} API with a custom \texttt{PortfolioEnv}. Specifically, at each trading day $t$ :
\begin{itemize}
  \item the \textbf{state} \(s_t\) concatenates flattened feature tensors
        \(\mathbf{x}_{t}\in R^{n_{\text{stocks}}\times n_{\text{feat}}}\)  
        and current portfolio weights \(\mathbf{w}_{t}\in\Delta^{n_{\text{stocks}}}\):
        \[
          s_t = \bigl[\operatorname{vec}(\mathbf{x}_{t}),\,\mathbf{w}_{t}\bigr];
        \]
        the \(n_{\text{feat}}=7\) features listed on Slide 2 capture return, momentum, sentiment, liquidity and volatility.
  \item the \textbf{action} \(a_t\in R^{n_{\text{stocks}}+1}\) is a set of raw logits that
        are \emph{projected} into valid long-only weights via softmax. 
  \item the \textbf{reward function} constructed is:
        \[
          r_t \;=\;
          \underbrace{\frac{V_{t+1}}{V_{t}}-1}_{\text{gross portfolio return}}
          \;-\;
          \underbrace{\text{turnover}_t\,c_{\text{tcost}}}_{\substack{\text{proportional}\\\text{transaction cost}}}
          \;-\;
          \underbrace{\text{short\_exposure}_t\,c_{\text{borrow}}}_{\text{borrow cost}},
        \]
\end{itemize}

\paragraph{Training-testing protocol:}
The training period of the reinforcement learning agent spans from 2018 to 2023. TD3 algorithm is fitted for 512 epochs using the 5-year in-sample data. Upon completion of training, the learned policy is frozen and evaluated on an out-of-sample period from January 2024 to January 2025. During this back-testing period, the trained actor is re-played in a separate environment with identical cost parameters to collect out-of-sample trading trajectories for performance assessment.

\vspace{1em}
\textbf{Execution Assumption:}
\begin{itemize}
    \item Orders are executed at the same day’s closing price (\textit{i.e.}, trades filled at the closing auction (\(\approx\) 4 p.m. ET).  )
    \item Long-only Strategy with initial investment of \$1 million, a level considered small enough to avoid exerting significant market impact. 
    \item Transaction cost is set conservatively at 10bps.
\end{itemize}

The caveats are the same as that applied in the rule-base strategy, except that the transaction cost for RL-driven strategy is set at a more conservative 10bps per trade.

\section{Results}
\subsection{Rule-Based Long-Short Strategy Performance Result}
The rule-based long-short strategy was tested in both the out-of-sample period from Jan 2024 to Jan 2025 and the full-sample period from Jan 2018 to Jan 2025.
\vspace{1em} 
\textbf{OOS Results}

The following table summarizes the key performance metrics for each sentiment weight:

\begin{table}[h!]
\centering
\resizebox{\textwidth}{!}{
\begin{tabular}{lccc}
\toprule
\textbf{Metric} & \textbf{Sentiment Weight = 0} & \textbf{0.5} & \textbf{1.0} \\
\midrule
Annualized Return (\%) & 20.14\% & 16.66\% & 15.55\% \\
Volatility (\%)        & 11.78\% & 11.70\% & 11.70\% \\
Sharpe Ratio           & 1.6146  & 1.3735  & 1.2916  \\
Sortino Ratio          & 0.1442  & 0.1220  & 0.1092  \\
Max Drawdown (\%)      & -5.63\% & -6.78\% & -7.66\% \\
\bottomrule
\end{tabular}
}
\caption{Performance Metrics of OOS Combined Strategy with Different Sentiment Weights}
\end{table}

We observe that all the long-short strategy has robust return profile in the OOS backtest. While increasing sentiment weight tends to slightly reduce return and Sharpe ratio, it also leads to higher drawdowns and lower downside protection as reflected in the Sortino ratio.

\begin{figure}[h!]
\centering
\begin{subfigure}[t]{0.48\textwidth}
    \centering
    \includegraphics[width=\linewidth]{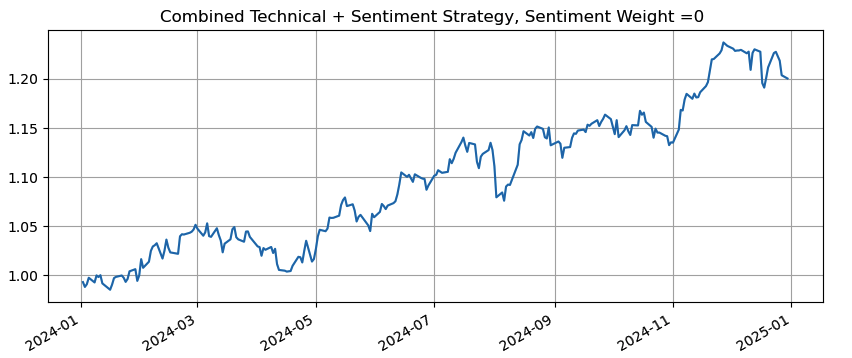}
    \caption{Sentiment Weight = 0}
\end{subfigure}
\hfill
\begin{subfigure}[t]{0.48\textwidth}
    \centering
    \includegraphics[width=\linewidth]{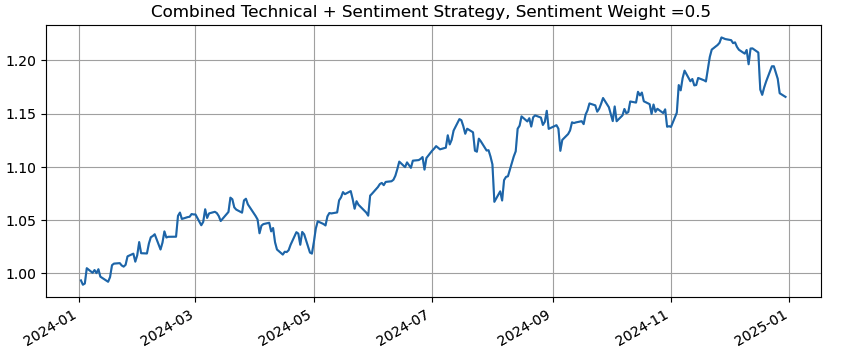}
    \caption{Sentiment Weight = 0.5}
\end{subfigure}

\vspace{0.5em}

\begin{subfigure}[t]{0.48\textwidth}
    \centering
    \includegraphics[width=\linewidth]{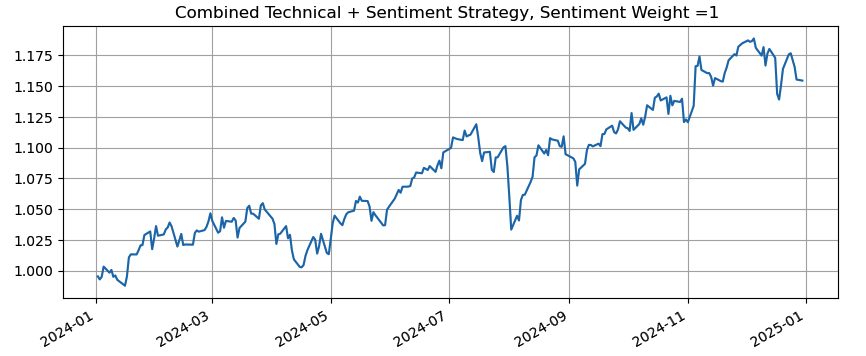}
    \caption{Sentiment Weight = 1}
\end{subfigure}

\caption{OOS Backtest NAV under Varying Sentiment Weights}
\label{fig:oos_backtest}

\end{figure}
\vspace{1em} 
\textbf{Fama-French Factor Decomposition for OOS Results}

To further examine the OOS return attributions, we regress the strategy return on the Fama-French 5 Factors to look at our strategy exposure. The regression result is shown below:
\begin{table}[h!]
\centering
\resizebox{\textwidth}{!}{
\begin{tabular}{lccc}
\toprule
\textbf{Factor} & \textbf{Sentiment = 0} & \textbf{Sentiment = 0.5} & \textbf{Sentiment = 1} \\
\midrule
Const            & 0.0006              & 0.0004              & 0.0004 \\
Market (mktrf)   & -0.0196              &  -0.0310              & -0.0206 \\
SMB              &  0.0791              &  0.0661                &  0.0199 \\
HML              &   0.0408            &   0.0636                &  0.0982 \\
RMW              & -0.0733               &  -0.0611              & -0.1236 \\
CMA              &  0.1166              &   0.1263               &  0.0972 \\
UMD              & 0.1118        &\textbf{0.1539$^{*}$}           & \textbf{0.1380$^{*}$} \\
\midrule
\textbf{R-squared} & 0.022 & 0.025 & 0.025 \\
\bottomrule
\end{tabular}
}
\caption{Factor Loadings from Fama-French 5-Factor + Momentum Model (OOS Strategy)}
\label{tab:ff_oos}
\end{table}

\begin{itemize}
    \item None of the factor loadings are statistically significant at conventional levels(0.05) across sentiment weights, implying few meaningful linear relationship between the strategy returns and standard risk premia.
    \item The absence of significant exposures suggests that the OOS strategy may be capturing alpha that is unrelated to market, size, value, profitability, investment, or momentum factors.
    \item The R-squared values are consistently low (approximately 2--3\%), indicating that the Fama-French 5-Factor + Momentum model has limited explanatory power for this strategy in the out-of-sample period.
    \item These findings imply that the strategy’s return drivers likely lie outside of conventional factor-based explanations, and may be rooted in alternative sources such as technical or sentiment-driven signals.
\end{itemize}

Since few coefficients are significant, we will later examine the full samplr return decomposition in the next part and propse several explanations for this result.


\vspace{1em} 
\textbf{Full Sample Results}

The following table summarizes the key performance metrics for each sentiment weight for the full sample period from 2018 to 2025:

\begin{table}[h!]
\centering
\resizebox{\textwidth}{!}{
\begin{tabular}{lccc}
\toprule
\textbf{Metric} & \textbf{Sentiment Weight = 0} & \textbf{0.5} & \textbf{1.0} \\
\midrule
Annualized Return (\%) & 13.84\% & 13.58\% & 13.25\% \\
Volatility (\%)        & 20.10\% & 20.14\% & 20.15\% \\
Sharpe Ratio           & 0.7475  & 0.7350  & 0.7203  \\
Sortino Ratio          & 0.0566  & 0.0561  & 0.0547  \\
Max Drawdown (\%)      & -37.32\% & -36.24\% & -35.49\% \\
\bottomrule
\end{tabular}
}
\caption{Performance Metrics of Combined Strategy with Different Sentiment Weights (2018–2025)}
\end{table}

We observe that the combined strategy maintains a relatively stable return profile across different sentiment weights over the full-sample period from 2018 to 2025. As sentiment weight increases, both annualized return and Sharpe ratio decrease slightly, while maximum drawdown narrows modestly. This suggests that a a higher sentiment weight may offer marginal risk reduction, but at the cost of a lower risk-adjusted return.

\newpage

\begin{figure}[h!]
\centering
\begin{subfigure}[t]{0.48\textwidth}
    \centering
    \includegraphics[width=\linewidth]{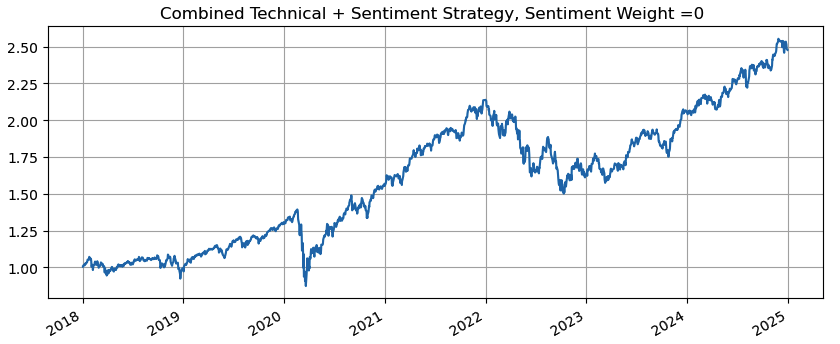}
    \caption{Sentiment Weight = 0}
\end{subfigure}
\hfill
\begin{subfigure}[t]{0.48\textwidth}
    \centering
    \includegraphics[width=\linewidth]{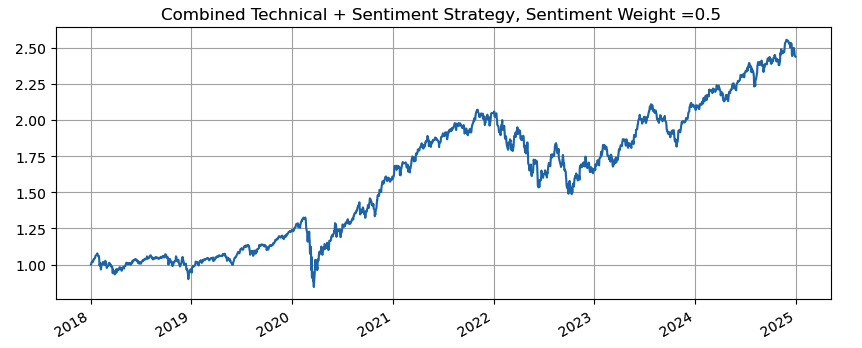}
    \caption{Sentiment Weight = 0.5}
\end{subfigure}

\vspace{0.5em}

\begin{subfigure}[t]{0.48\textwidth}
    \centering
    \includegraphics[width=\linewidth]{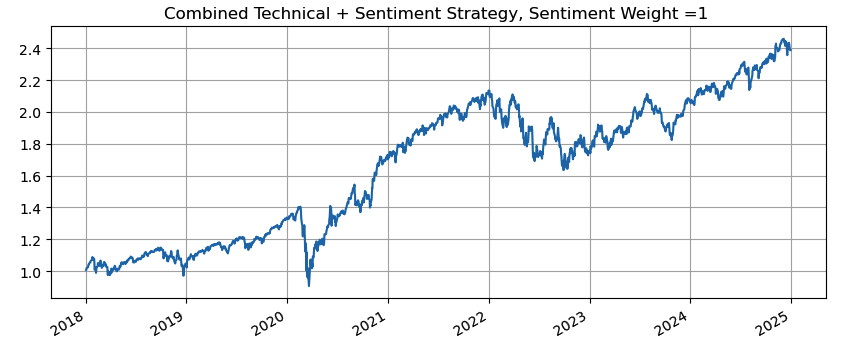}
    \caption{Sentiment Weight = 1}
\end{subfigure}

\caption{Full Sample Backtest NAV under Varying Sentiment Weights}
\label{fig:full_backtest_all}
\end{figure}

\textbf{Fama-French Factor Decomposition for Full Sample Results}

Similarly, we did the return attribution analysis on the full sample period and get the FF5 Factor exposure results as follows:
\begin{table}[h!]
\centering
\resizebox{\textwidth}{!}{
\begin{tabular}{lccc}
\toprule
\textbf{Factor} & \textbf{Sentiment = 0} & \textbf{Sentiment = 0.5} & \textbf{Sentiment = 1} \\
\midrule
Const            & 0.0006$^{*}$              & 0.0006$^{*}$              & 0.0006$^{*}$ \\
Market (mktrf)   & \textbf{-0.1675$^{***}$}  & \textbf{-0.1648$^{***}$}  & \textbf{-0.1841$^{***}$} \\
SMB              & \textbf{0.1763$^{***}$}   & \textbf{0.1776$^{***}$}   & \textbf{0.1711$^{***}$} \\
HML              & \textbf{-0.1342$^{***}$}  & \textbf{-0.1524$^{***}$}  & \textbf{-0.1229$^{***}$} \\
RMW              & -0.0857                   & 0.0687                    & 0.0528 \\
CMA              & -0.0372                   & -0.0253                   & -0.0253 \\
UMD              & -0.0416                   & \textbf{-0.0507$^{*}$}    & \textbf{-0.0559$^{**}$} \\
\midrule
\textbf{R-squared} & 0.033 & 0.034 & 0.038 \\
\bottomrule
\end{tabular}
}
\caption{Fama-French 5-Factor + Momentum Regression: Full Sample Loadings Across Sentiment Weights}
\label{tab:ff_full_all}
\end{table}

\begin{itemize}
    \item The strategy exhibits a consistently significant and negative loading on the market factor (MKT), suggesting a contrarian beta profile across all sentiment specifications.
    \item Exposure to the size factor (SMB) is positive and highly significant throughout, indicating a persistent tilt toward small-cap names.
    \item The strategy also shows significantly negative exposure to the value factor (HML), implying a preference for growth stocks over value.
    \item The momentum factor (UMD) becomes increasingly significant as sentiment weight increases. For sentiment = 1, the strategy loads negatively and significantly on UMD, reflecting potential anti-momentum behavior when sentiment signals dominate.
    \item Other factors such as RMW and CMA do not exhibit statistically meaningful influence, indicating limited sensitivity to profitability or investment style.
    \item Overall \( R^2 \) values are low (3–4\%), suggesting that traditional factor models explain only a modest portion of the strategy’s returns, reinforcing its potential uniqueness or alpha beyond standard risk premia.
\end{itemize}
\vspace{1em} 
\textbf{Potential Explanation of the OOS Return Decomposition Insignificance}
\begin{itemize}
    \item \textbf{Market concentration and regime shift}: In 2024, the U.S. equity market experienced significant concentration, with the "Magnificent 7" stocks comprising a substantial portion of the S\&P 500's market capitalization. This concentration may have altered traditional factor exposures, affecting the performance of the strategy and its relationship with standard risk factors.

    \item \textbf{Sample size limitations}: The OOS period encompasses only one year (2024), providing a limited number of observations for regression analysis. This smaller sample size reduces statistical power, making it more challenging to detect significant relationships between the strategy returns and risk factors.

    \item \textbf{Structural changes in market dynamics}: The unique economic and political events of 2024 may have led to structural changes in market dynamics, rendering traditional factor models less effective in capturing the drivers of strategy returns during this period.

\end{itemize}
\vspace{1em} 
\textbf{Transaction Cost Analysis}

The rule-based strategy was tested under the assumption that no transaction cost incurred during the execution process. As a result, there is an significant alpha as showned in the full sample return decomposition. To test the strategy robustness, a transaction cost of 5 basis points is applied to each trade in the following backtests, which is high and conservative for the large-cap stocks in the portfolio, mitigating potential overstatement of performance results. The results are reported as follows:

\newpage

\begin{table}[h!]
\centering
\resizebox{\textwidth}{!}{
\begin{tabular}{lccc}
\toprule
\textbf{Metric} & \textbf{Sentiment Weight = 0} & \textbf{0.5} & \textbf{1.0} \\
\midrule
Annualized Return (\%) & 3.66\%  & 0.13\%  & 1.22\%  \\
Volatility (\%)        & 20.10\% & 20.14\% & 20.15\% \\
Sharpe Ratio           & 0.2806  & 0.1075  & 0.1614  \\
Sortino Ratio          & 0.0214  & 0.0083  & 0.0124  \\
Max Drawdown (\%)      & -37.86\% & -40.20\% & -36.25\% \\
\bottomrule
\end{tabular}
}
\caption{Full Sample (2018–2025) Performance of Combined Strategy, Transaction Cost = 5bps}
\label{tab:full_sample_sentiment}
\end{table}

\begin{figure}[h!]
\centering
\begin{subfigure}[t]{0.48\textwidth}
    \centering
    \includegraphics[width=\linewidth]{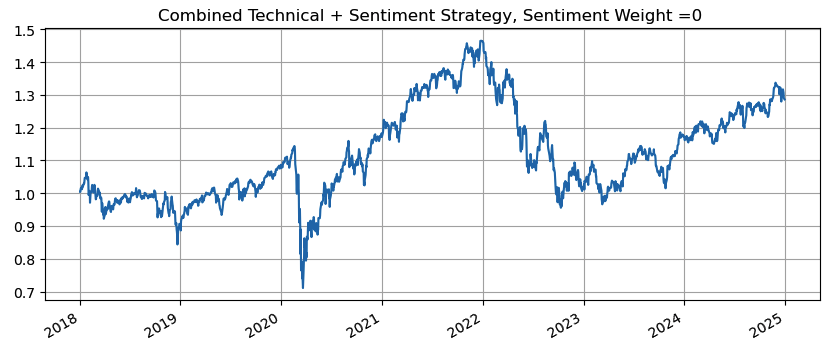}
    
    \caption{Sentiment Weight = 0}
\end{subfigure}
\hfill
\begin{subfigure}[t]{0.48\textwidth}
    \centering
    \includegraphics[width=\linewidth]{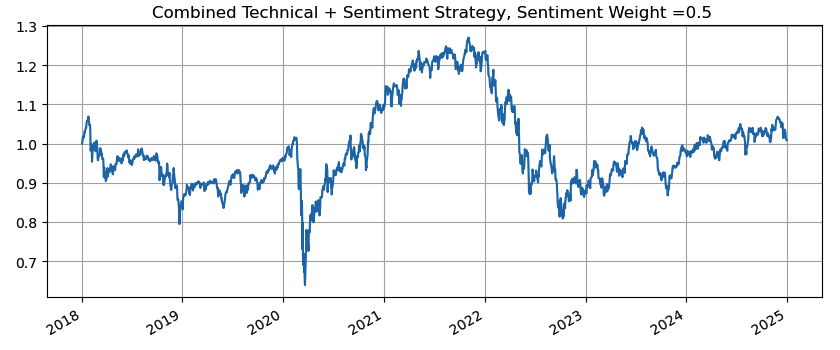}
    \caption{Sentiment Weight = 0.5}
\end{subfigure}

\vspace{0.5em}

\begin{subfigure}[t]{0.48\textwidth}
    \centering
    \includegraphics[width=\linewidth]{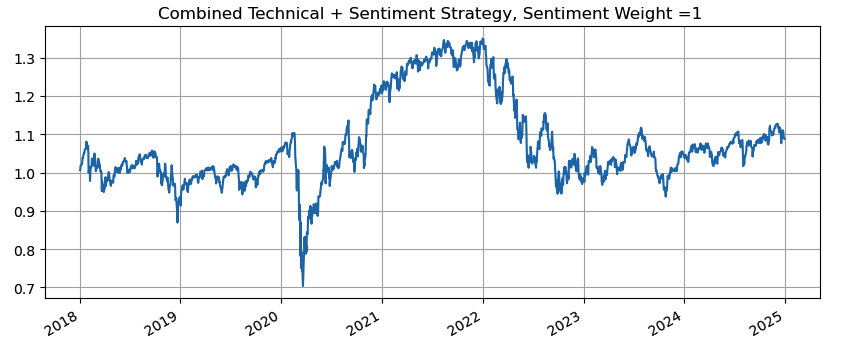}
    \caption{Sentiment Weight = 1}
\end{subfigure}

\caption{Full Sample Backtest NAV under Varying Sentiment Weights, Transaction Cost = 5bps}
\label{fig:full_backtest_all}
\end{figure}
\textbf{Impact of Transaction Costs on Strategy Performance}

Comparing the two full sample performance tables, we observe a clear deterioration in annualized return once a 5bps transaction cost is applied. For example, under sentiment weight = 0, the annualized return drops from 13.84\% to 3.66\%, representing a more than 10\% absolute reduction. Similar erosions are observed across other sentiment weights, with Sharpe ratios also significantly declining. The effect is more significant in higher turnover regimes where strategy is assigned with higher sentiment weights, and this explains the more substantial drop in return for sentiment weight = 0.5 and 1.0. Despite volatility remaining stable, the deterioration in both absolute and risk-adjusted returns underscores the importance of cost-aware strategy design in practical implementation.

\subsection{RL-driven Strategy Performance Results}
\label{subsec:rl_results}

\begin{figure}[h!]
  \centering
  \includegraphics[width=0.9\linewidth]{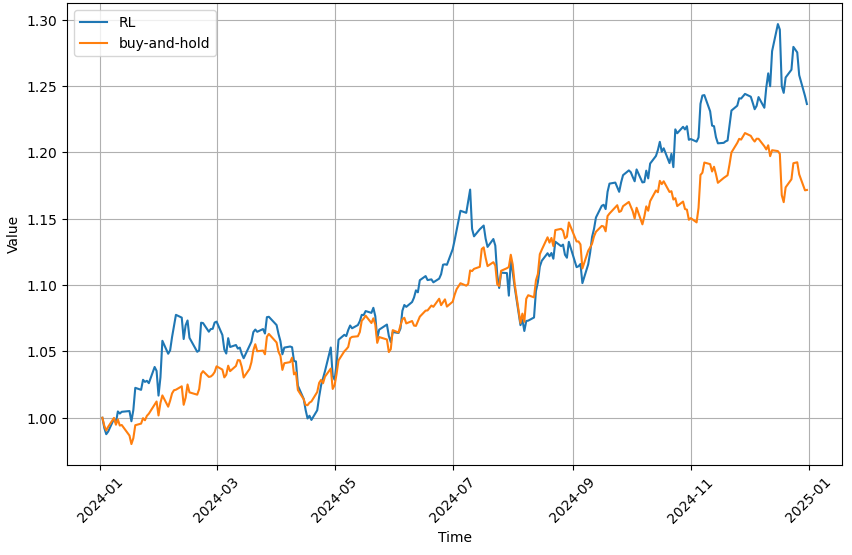}
  \caption{OOS portfolio growth: TD3 vs.\ Buy \& Hold (Jan 2024 –Jan 2025)}
  \label{fig:rl_nav}
\end{figure}

Figure \ref{fig:rl_nav} plots the net asset value (NAV) of the TD3 policy against the long-only buy-and-hold benchmark in the 2024 out-of-sample window. Table \ref{tab:rl_metrics} summarizes the key statistics for the strategy.

\begin{table}[h!]
\centering
\begin{tabular}{lcc}
\toprule
\textbf{Metric} & \textbf{TD3 Policy} & \textbf{Buy \& Hold} \\
\midrule
Annualised return & 23.65\% & 17.17\% \\
Annualised volatility & 13.46\% & 10.06\% \\
Sharpe ratio & 1.38 & 1.20 \\
Sortino ratio & 1.96 & 1.59 \\
Portfolio turnover & 52.3\% & 0.0\% \\
Max drawdown & –9.09\% & –5.06\% \\
\bottomrule
\end{tabular}
\caption{Out-of-sample performance metrics (Jan 2024 – Jan 2025)}
\label{tab:rl_metrics}
\end{table}

\textbf{FF-5 Factor Decomposition:}
\vspace{1em}

A daily-company-level regression of the excess returns of TD3 in the Fama-French 5-factor + momentum model yields \(R^{2}=0.65\) and an insignificant intercept (\(p=0.93\)), indicating that the market beta explains most of the variation, while the idiosyncratic alpha is statistically non-existant once transaction costs are considered. The significant factor loadings are \(\beta_{\text{mktrf}}=0.91\), \(\beta_{\text{smb}}=-0.17\) and \(\beta_{\text{umd}}=-0.14\), implying a defensive, large-cap, anti-momentum stance—consistent with the policy holding winners longer and rotating into laggards after momentum reversals.

\section{Conclusion and Discussion}

In this paper, we leverage domain-specific large language model to perform sentiment analysis on financial news articles and integrate the sentiment signals with traditional technical indicators into dynamic quantitative trading strategies. Our approaches include both a conventional rule-base integration and a more complex reinforcement learning-driven framework. Specifically, the comparison between the two strategies shed light on the effectiveness of the sentiment signal itself and the reinforcement learning algorithm.
\newline

The results from both the rule-based and RL-driven strategies indicate that the sentiment signal extracted from Thomson Reuters and generated by FinGPT provides some added value over strategies based solely on technical indicators. In the rule-based long-short strategy, portfolios with positive sentiment weights exhibit lower maximum drawdowns, although they tend to yield lower annual returns and Sharpe ratios compared to purely technical strategies. In the rule-base long-only strategy shown in the appendix, portfolio with positive sentiment weights exhibit better performance. In the RL-driven strategy, the portfolio exceeds the benchmark buy-and-hold portfolio in annual return, Sharpe ratio and Sortino ratio when incorporating the sentiment signals. 
\newline

Furthermore, the RL agent delivers a Sharpe ratio of 1.38 despite a 52 \% annual turnover even after imposing a conservatively high-level transaction cost of 10bps, demonstrating that dynamic re-allocation outweighs trading frictions.In a realistic trading environment, the transaction costs for these mega-cap stocks are expected to be much lower, which would likely result in more favorable returns for the RL-driven strategy. This paper shows that reinforcement learning is to some degree an effective and promising way to integrate traditional factors with signals from unstructured data such as LLM-generated sentiment scores.
\newline

In particular, we find that both rule-base strategy with positive sentiment weight and RL-driven strategy both exhibit high turnover ratio. In the rule-based strategy, portfolio returns deteriorate significantly after accounting for transaction costs. The RL-driven strategy also delivers a 52\% annual turnover ratio, albeit much lower than that in the rule-base portfolio. This indicates a key characteristic of sentiment signals - the rich and frequently updated news flow calls for more frequent portfolio re-balancing.
\newline

\paragraph{Limitations and Future Research} First, due to limited GPU resources, our current analysis is restricted to a smaller universe of stocks. We focus on mega-cap stocks, as they tend to receive more consistent news coverage, which is an essential consideration given our limited access to diverse news sources. Expanding the stock universe to include smaller-cap stocks would require additional data sources to ensure sufficient sentiment signal quality. Second, the RL-driven strategy’s performance is sensitive to the cash bias introduced by the softmax projection. Future research will explore alternative portfolio constraints, including risk budgeting and variance-penalized reward functions, to better control tracking error and improve allocation stability.

\section{Appendix}

\textbf{Full Sample Result for Long-Only Strategy}

The following table summarizes the key performance metrics for each sentiment weight for the full sample period of long-only strategy from 2018 to 2025:

\begin{figure}
    \centering
    \includegraphics[width=1\linewidth]{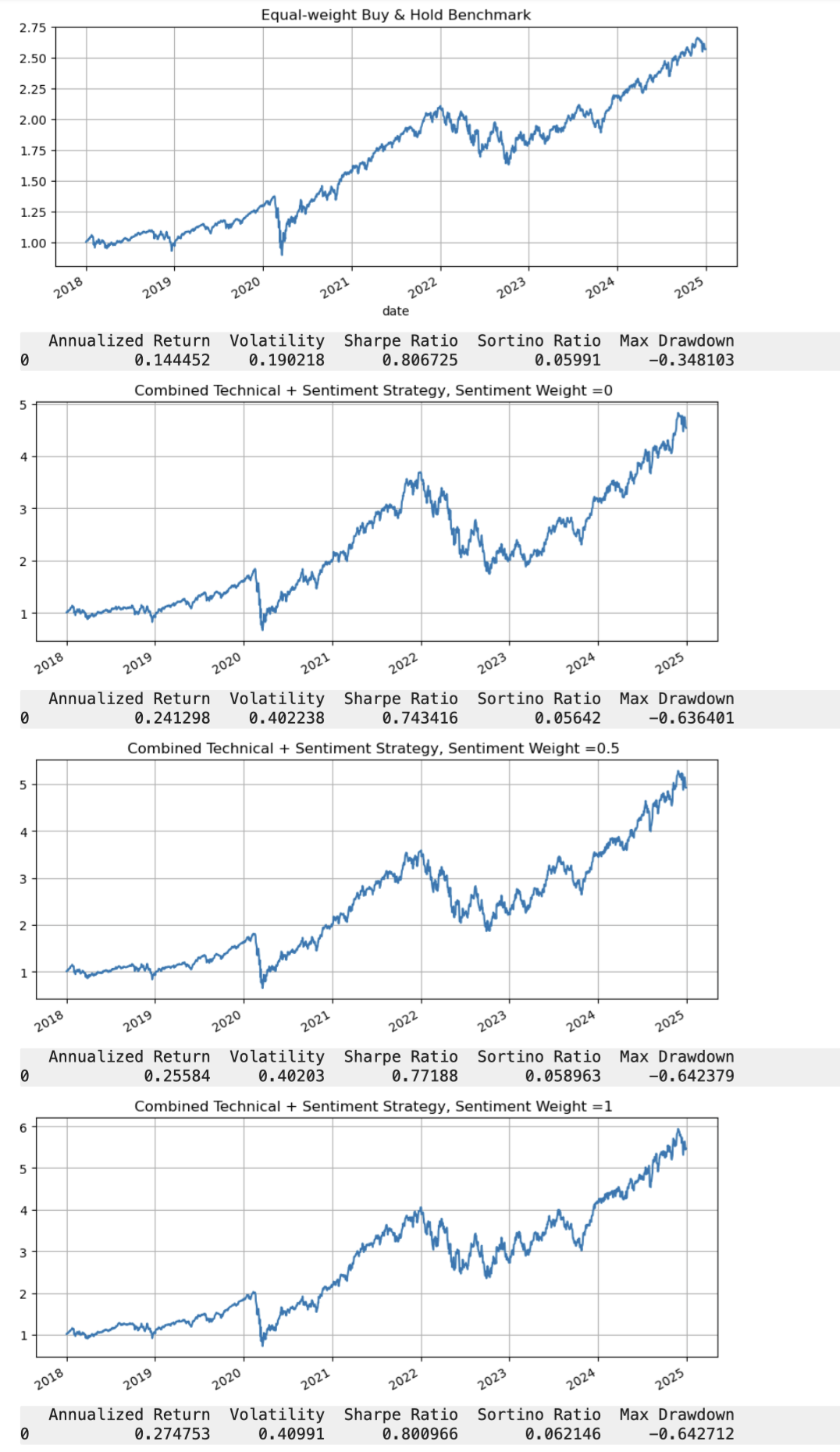}
    \caption{Long-only Full Sample Results with Zero Transaction Costs}
    \label{fig:enter-label}
\end{figure}

\clearpage

\end{document}